# 3D-localization microscopy and tracking of $F_oF_1$-ATP synthases in living bacteria

Anja Renz[a], Marc Renz[a], Diana Klütsch[b], Gabriele Deckers-Hebestreit[b], Michael Börsch*[a,c,d]

[a]Single Molecule Microscopy Group, Jena University Hospital, Friedrich Schiller University Jena, Nonnenplan 2 - 4, D-07743 Jena, Germany; [b]Department of Microbiology, University of Osnabrück, Barbarastrasse 11, D-49076 Osnabrück, Germany; [c]Jena Center for Soft Matter (JCSM); [d]Abbe Center of Photonics (ACP), Jena, Germany

## ABSTRACT

$F_oF_1$-ATP synthases are membrane-embedded protein machines that catalyze the synthesis of adenosine triphosphate. Using photoactivation-based localization microscopy (PALM) in TIR-illumination as well as structured illumination microscopy (SIM), we explore the spatial distribution and track single $F_oF_1$-ATP synthases in living *E. coli* cells under physiological conditions at different temperatures. For quantitative diffusion analysis by mean-squared-displacement measurements, the limited size of the observation area in the membrane with its significant membrane curvature has to be considered. Therefore, we applied a 'sliding observation window' approach (M. Renz *et al., Proc. SPIE* 8225, 2012) and obtained the one-dimensional diffusion coefficient of $F_oF_1$-ATP synthase diffusing on the long axis in living *E. coli* cells.

## 1. INTRODUCTION

Optical superresolution techniques have had a rapid and astonishing progression during recent years since 2006[1-4], culminating in the Nobel Prize for chemistry in 2014. Methods related to PALM (photoactivated localization microscopy) and STORM (stochastic optical reconstruction microscopy) are widely used to address a range of biological or biomedical questions. We used these techniques to study the distribution and motility of the membrane enzyme $F_oF_1$-ATP synthase in living *Escherichia coli* cells[5]. The main question pertains to the formation of functional membrane protein supercomplexes of different parts of the oxidative phosphorylation process (OXPHOS). OXPHOS is under aerobic conditions the most common way of ATP production. All participating enzymes, often numbered complex I to V, with V being the $F_oF_1$-ATP synthase, are located in the cytoplasmic membrane of bacteria, like *E. coli*, in the thylakoid membranes of chloroplasts or in the inner membrane of mitochondria. For fast and efficient generation of the biochemical 'energy currency' ATP, direct contact between different OXPHOS enzymes in mitochondria has been proposed[6].

Newer models propose the presence of OXPHOS supercomplexes also for bacteria[7-12], located as patches in the plasma membrane and related to specific lipid compositions[12]. This view is supported by new insight into the composition of bacterial membranes[13] with the bacterial cytoskeleton MreB[14-16] being involved in membrane organization[17]. It seems that the long held view of a 'lipid mosaic'[18] is challenged by a model of 'lipid rafts' or 'lipid (nano)domains'[19], in which specific lipids and lipid mixtures condense into segregated rafts or patches involving flotellins as organizing element[13, 17]. Whether these rafts provide the means for the formation of protein supercomplexes or whether protein interactions are the cause for the presence of membrane microcompartmentalization is still under dispute; as is also the nature, size or existence of rafts themselves.

Superresolution microscopy techniques like PALM, STORM, STED or SIM allow to optically probe nanometer-sized regions[20] proposed for this kind of microcompartments[12]. Here, we have investigated the spatial distribution and analyzed the diffusion properties of $F_oF_1$-ATP synthases in living *E.coli* cells by single enzyme tracking (SPT) using either EGFP, paGFP or mEOS3.2 as fluorescent markers fused to the C terminus of the membrane-integral $F_o$ subunit *a*.

.................................................................................................................................................

*michael.boersch@med.uni-jena; phone +49 3641933745; fax +49 3641933750; http://www.m-boersch.org

## 2. EXPERIMENTAL PROCEDURES

**2.1 Mutant constructions and their expression levels**

Mutant strain HW3 (*atpB::mEos3.2*) expressing chromosomally $F_oF_1$-ATP synthase with photoactivatable, monomeric mEos3.2 fused to the C terminus of subunit *a* was generated from *E. coli* strain K-12 in two steps by the λ Red system(21). First, mutant strain EB4 (*ΔatpBE::aphA; phenotype Succ⁻ Kan*$^R$) was generated by deleting genes *atpBE* of the *atp* operon by simultaneously inserting a kanamycin resistance cassette (Kan$^R$) for selection. Due to the deletion, cells have a defect in oxidative phosphorylation and cannot grow on succinate (Succ⁻). To obtain the corresponding PCR product primers 5'-*GGTGCTGGTGGTTCAGATACTGGCACCGGC*-TGTAATTAACAACAAAGGGGTGTAGGCTGGAGCTGCTTC-3' and 5'-*GCAGAGGAATCAGATCAGGTTGACGCG*-CTGCGCCTTCCAGGAATTTACCCATATGAATATCCTCCTTAG-3' were used. Deoxynucleotides corresponding to the flanking sites of the *atp* operon are in italics, the priming sites of the template plasmid pKD4(21) carrying the resistance cassette are in standard. The *atpBE* deletion, which starts prior to the stop codon of *atpI* and ends at codon 32 of the *atpE* gene, was verified by colony PCR. In the second step, the kanamycin resistance cassette was exchanged using for homologous recombination the 2631 bp ScaI/AseI fragment of pHW3ok and cells expressing again an ATP synthase functional in oxidative phosphorylation were selected by growth on minimal medium with succinate(22). As a result, strain HW3 shows a Succ⁺ Kan$^S$ phenotype during growth and carries an *atp* operon with an *atpB::mEos3.2* fusion gene as verified by colony PCR as well as DNA sequencing.

For construction of plasmids pHW3ok and pBH107, a BamHI restriction site (encoding a Gly-Ser spacer) and the gene encoding mEos3.2 (pHW3ok) or paGFP (pBH107) were inserted prior to the stop codon into *atpB* (encoding subunit *a*) using a two-step PCR overlap extension method. In the first step, three different PCR products were generated using (*i*) primers 5'-GTGACTGGTGAGTACTCAACCAAGTC-3' and 5'-CATGTCTGGCTTAATCGCACTCAT**GGATCC***AT-GATCTTCAGACGC*-3' with pSD166(23) as template, (*ii*) primers 5'-*GCGTCTGAAGATCAT***GGATCC**ATGAGTGCG-ATTAAGCCAGACATG-3' and 5'-*CGTAGTAGTGTTGGTAAATTA*TCGTCTGGCATTGTCAGGCAATCC-3' with pmEos3.2-N1(24) for pHW3ok or and primers 5'-GCGTCTGAAGATCATGGATCCATGGTGTCTAAGGGCGAAG-AGC-3' and 5'-CGTAGTAGTGTTGGTAAATTAGAGGGATCCTCATCTGTGCCCC-3' with pSems-paGFP (kindly provided by Jakob Piehler, Osnabrück) for pBH107, and (*iii*) primers 5'-*GGATTGCCTGACAATGCCAGACGA*TAA-TTTACCAACACTACTACG-3' and 5'-*GCGTTAAGATTCACAGCACAATGCC*-3' with pBWU13(25). Deoxynucleotides corresponding to the *atp* operon are marked in italics, those of the mEos3.2 gene or the paGFP gene are underlined, and the spacer is in bold letters and the vector part in standard. The three PCR products were annealed and the first and last primer used for the second amplification step. After restriction of the PCR product with ScaI and BsrGI, the resulting 2569 bp fragment was introduced into correspondingly restricted pBH9 (a pBWU13 derivative missing the restriction sites AseI in *bla* and BsrGI in *atpG*) generating plasmid pHW3ok or pBH107. The fusion genes were verified by DNA sequencing through the ligation sites.

Cloning of the EGFP fusion to the C terminus of subunit *a* of $F_oF_1$-ATP synthase has been published previously(23).

To estimate the relative $F_oF_1$-ATP synthase concentration in cell membranes, *E. coli* cells were grown overnight at first in LB (lysogeny broth) medium and subsequently grown in 500 mL of fermenter medium(23). About 1 g cells were harvested and washed. Cells were broken using a cell disruptor, and about 0.5 g cell membranes were recovered after ultra-centrifugation. Protein quantification was performed using the amido black or BCA assay (Pierce). Membranes were loaded on SDS-PAGE and stained with Coomassie Blue(26) or analyzed by immunoblotting using antibodies raised against individual $F_oF_1$ subunits(27). ATP hydrolysis activities as well as its N,N'-dicyclohexylcarbodiimide sensitivities were performed as described(28).

**2.2 Sample preparation for microscopy**

Cells were grown overnight in LB medium and subsequently diluted 200 times in minimal medium M9 with glucose as carbon and energy source to reduce autofluorescence(29). After 3 hours of aerobic growth at 37°C, cells were washed two times in PBS buffer to remove the growth medium. *E. coli* cells were attached to poly-lysine-coated cover glass in a home-built imaging chamber(5). To ensure a flat orientation of the cells on the surface and to get rid of diffusing *E.coli* cells, sample chambers were rinsed with either PBS buffer or M9.

For PALM imaging, anionic fluorescent beads with 0.1 μm diameter (TetraSpeck, Molecular Probes) were diluted 1:500 and added to the chamber after *E. coli* cells. Beads were used as fiducial marks to correct for sample drift and to control the position of the focus plane. Chambers were sealed with nail polish to prevent drying-out immediately before microscopy measurements.

**2.3 Microscope setup**

Bacteria were imaged on an inverted Nikon microscope (N-SIM / N-STORM) with a CFI Apo TIRF 100x oil objective lens (N.A. 1.49). Fluorescence images were recorded by an Andor iXon DU897 EMCCD camera controlled by the Nikon measurement software. The camera was attached *via* a turret which allows adjustment of the magnification depending on the measurement mode. We used either two-dimensional PALM (2D-PALM) for localization in flat parts of the bacterial membrane, or three-dimensional PALM (3D-PALM) using an additional astigmatic lens. For SIM, an additional 2.5x magnification lens was inserted before the EMCCD. SIM measurements were performed with a pixel size of 60 nm and five grid positions for each of the three grid orientations. 3D-PALM imaging was accomplished with a 100 nm pixel size by inserting the 1.5x magnification lens, and SPT measurements were carried out with a 160 nm pixel size.

Laser excitation was provided by fiber-coupled continuous-wave lasers at 405 nm, 488 nm and 561 nm. Laser light was directed to the microscope objective by a quad line beam splitter zt405/488/561/640rpc-TIRF (F73-410, AHF Analysentechnik). Different band pass filters (AHF) were used to select fluorescence: ET 525/50 for EGFP and paGFP, or ET 600/50 for mEOS3.2. For SIM excitation, the 100xEx V-R grating block was used that allows excitation at different wavelengths from 488 nm up to 640 nm. SIM measurements were carried out in widefield illumination. For localization microscopy imaging, widefield, oblique angle or TIR illumination was used. For SPT measurements only TIR illumination was applied. Samples were positioned by a piezo stage in combination with a motorized x-y linear stage. The piezo z-positioner was used for 3D calibration in 3D-PALM, and for acquiring z-stacks using SIM. The Nikon Perfect Focus System ensured the stability of our z-positioning during all measurements.

3D localization was obtained by the astigmatism method. Calibration of Gaussian width in x and y as a function of z position was generated using the piezo stage insert. 3D-PALM was calibrated with 200x dilution of 0.1 μm TetraSpeck fluorescent beads in eight-well Lab-Tek II chambered cover glass (Nunc). Calibration steps included 20 frames in focus, followed by repositioning the piezo at -800 nm in z, then 10 nm increments of z-steps until +800 nm is reached and finally again 20 images in focus. Each objective and each laser line was calibrated separately, and calibration information was stored. 3D-PALM could cover maximal range of 800 nm in height after calibration.

For SIM measurements EMCCD exposure times were set to 200 ms, and we performed a 1 μm z-stack composed of 11 steps, each 0.1 μm apart. Using paGFP for PALM imaging, each 'activation frame' was followed by three 'reporter frames', and in total 40.000 frames were recorded. Exposure times were 30 ms per frame, and laser intensities were set to 33 W/cm$^2$ at 405 nm ('activation') and 855 W/cm$^2$ at 488 nm. During the activation frame, no image data were collected; only fluorescence information in the second and third frame after activation was used for PALM imaging analysis. 5000 images were collected for 2D-PALM imaging and between 15.000 and 20.000 for 3D-PALM imaging using mEOS3.2. Exposure times were set to 30 ms for 2D- and to 40 ms for 3D-PALM measurements using mEOS3.2. Laser intensities were maximal 0.5 W/cm$^2$ at 405 nm when needed, and maximal 260 W/cm$^2$ at 561 nm for 2D-PALM. Laser intensities were set to 112 W/cm$^2$ at 405 nm and 260 W/cm$^2$ at 561 nm. Continuous activation was used for 3D-PALM.

For SPT measurements series of several thousand images were collected (without the astigmatic lens inserted) with EMCCD exposure time of 30 ms. Laser intensities were 0.5 W/cm$^2$ at 405 nm when needed, and 260 W/cm$^2$ at 561 nm. EM-gain of the camera was set to 280 for SIM and to 250-300 for PALM imaging and SPT. Image recording was controlled by Nikon NIS-Elements AR 4.13.04 64-bit software. Processing of high-resolution images was performed in the Nikon NIS-Elements AR Analysis 4.20.00 64-bit software. For SPT data ImageJ and Matlab scripts were used as described below.

A Tokai Hit Incubation System was mounted onto the piezo stage to provide a sample temperature of 37°C. The Tokai Hit System also included an objective heating element to keep the temperature stable.

## 3. RESULTS

**3.1 Expression level and activities of chromosomally and plasmid-encoded fluorescent $F_oF_1$-ATP synthases**

To generate a fluorescently labeled $F_oF_1$-ATP synthase for superlocalization microscopy, fluorescent proteins mEos3.2(24), paGFP(30) and EGFP(23), respectively, were fused to the C terminus of the membrane-integral $F_o$ subunit *a* separated by a Gly-Ser linker. To obtain expression levels comparable to wild type cells, *E. coli* strain HW3 was generated from K-12 in two steps by allelic replacement of the chromosomal *atpB* with the fusion gene *atpB::mEos3.2* as described above. The selection of HW3 cells by growth on minimal medium with the non-fermentative succinate as sole carbon and energy source confirms the presence of a $F_oF_1$-ATP synthase functional in oxidative phosphorylation. An immunoblot analysis with subunit *a*-specific antibodies (Fig. 1, lane A) showed that a stable fusion protein with an apparent molecular mass of approx. 55 kDa is present in inverted membrane vesicles of HW3 instead of subunit *a* present in wild type cells. ATP hydrolysis activities of inverted membrane vesicles of HW3 are reduced by a factor of two compared to wild type (K-12: 1.0 µmol $P_i$·min$^{-1}$·mg$^{-1}$; HW3: 0.4 µmol $P_i$·min$^{-1}$·mg$^{-1}$). Nevertheless, the *N,N'*-dicyclohexylcarbodiimide inhibitory sensitivity of ATP hydrolysis activity observed was within the same range (63-65%) indicating that the fusion of mEos3.2 to subunit *a* has no influence on the tight coupling between the proton translocation in $F_o$ and the catalytic centers in $F_1$. Furthermore, the detection of $F_o$ subunit *b* and $F_1$ subunit β in an additional immunoblot (Fig. 1, lane B) revealed a comparable reduction of both subunits in HW3 supporting the view that the reduction in ATP hydrolysis activity is a result of the manipulation within the *atp* operon leading in general to a lower expression of the individual $F_oF_1$ subunits. Indeed, an observation that has been made in several cases for changes in close vicinity to the translational initiation region of *atpE*, the gene located downstream of *atpB* (G. Deckers-Hebestreit, personal observation). In summary, the results obtained demonstrate that the function of $F_oF_1$-ATP synthase is not changed by the fusion of mEos3.2 to $F_o$ subunit *a*.

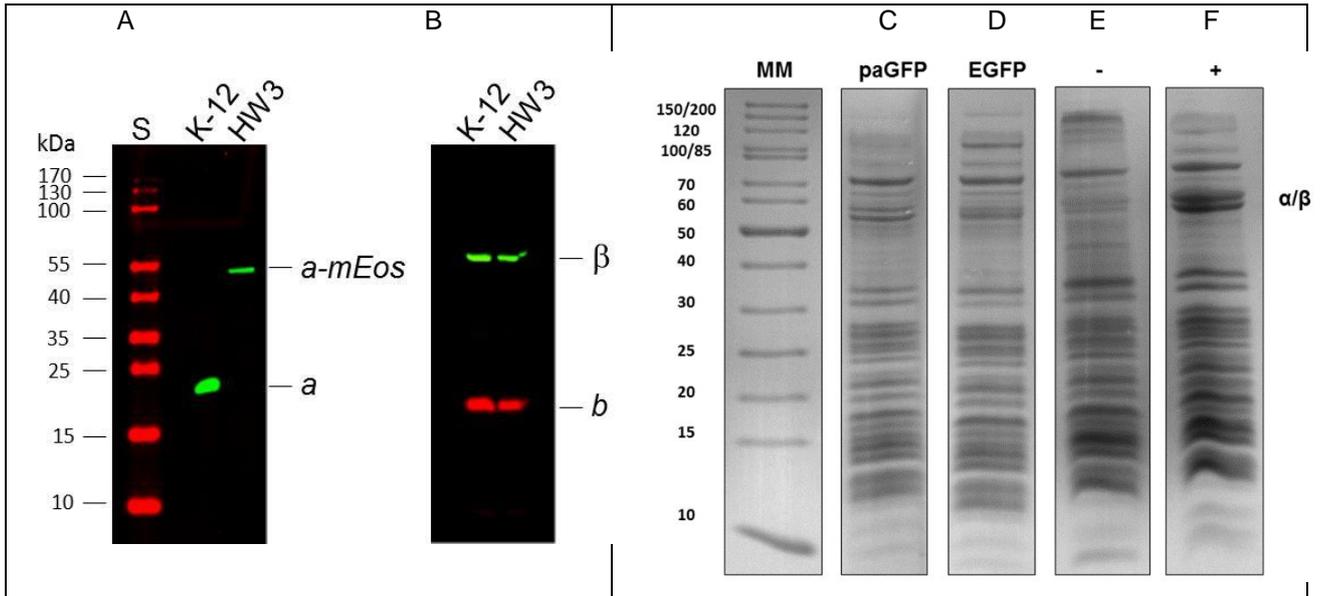

**Figure 1:** Lane **A** and **B,** Immunoblot analysis (20 µg of protein per lane) of chromosomally encoded $F_oF_1$-ATP synthase with fusion of fluorescent protein mEOS3.2 to the C terminus of subunit *a* (membranes of *E. coli* strain HW3) in comparison to membranes of the *E. coli* wild type strain K-12. Lane **C-F**, Coomassie-Blue stained (40 µg of protein per lane) SDS-PAGE of *E. coli atp* deletion RA1 transformed with plasmid pBH107 (**C**) or pSD166 (**D**) encoding $F_oF_1$-ATP synthases with fusion of paGFP (**C**) or EGFP (**D**) to subunit *a*. Membranes of *E. coli* strain RA1 without $F_oF_1$ (**E**) and transformed with plasmid pBWU13 expressing 'wild type' $F_oF_1$.

Fig. 1 shows a SDS-PAGE of *E. coli* membranes expressing plasmid-encoded $F_oF_1$-ATP synthases with paGFP (lane C) or EFGP (lane D) fusions to subunit *a* in an *atp* deletion background (*E. coli* strain RA1(31)). Lane E shows membranes of strain RA1 without $F_oF_1$-ATP synthase as the negative control, and in lane F with 'wild type $F_oF_1$' as a positive control. The relative amount of subunits α and β of $F_oF_1$-ATP synthases indicates that the expression level for the fusion constructs of $F_oF_1$-ATP synthase with EGFP and paGFP on subunit *a* is significantly decreased. Therefore, although the *atp* operon is plasmid-encoded, these transformed *E. coli* cells are ideal candidates for fluorescence microscopy. Even in *E. coli* cells expressing the $F_oF_1$-ATP synthase chromosomally several copies of the *atp* operon are at hand(32) due to the presence of the *atp* operon next to the origin of replication on the chromosome(33) and, furthermore, due to the presence of multiple replication forks in rapidly growing *E. coli* cells.

## 3.2 SIM Imaging

We used the EGFP fusion to subunit *a* of *E. coli* $F_oF_1$-ATP synthase encoded by RA1/pSD166 for SIM imaging. **Figure 2** shows the reconstructed SIM image of several *E. coli* cells as the middle layer in a z-stack at a height of 500 nm above the cover glass. Clear localization of $F_oF_1$-ATP synthases in the plasma membrane was possible. Intensity profiles perpendicular to the cell membranes yielded ~200 nm full-width-half-maximum. Because SIM required recording of 15 sub-images for each z layer, each reconstructed image integrates fluorescence over 3 seconds. Motion blur of the labeled enzymes at 22°C might cause the apparently uniform intensity distribution as observed in many of the *E. coli* cells.

However, other cells showed a few brighter spots in the membrane. Note, that no evidence was found for the presence for inclusion bodies at the cellular poles also indicating a rather moderate expression of $F_oF_1$-ATP synthase.

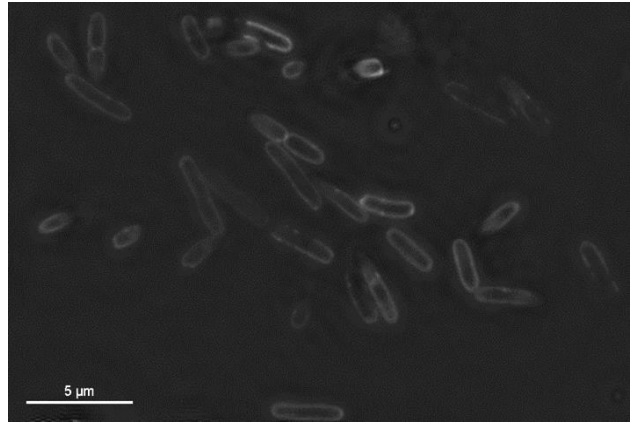

**Figure 2**: SIM image of plasmid-encoded $F_oF_1$-ATP synthases with EGFP fusion to subunit *a* in *E. coli*. We used the Nikon '3D-SIM' mode for each of the 11 layers in the z-stack. SIM image plane was located 500 nm above the cover glass. Laser excitation with 488 nm, scale bar is 5 μm.

### 3.3 PALM Imaging

Next, we investigated the spatial distribution of $F_oF_1$-ATP synthases with 2D- and 3D-PALM. Similar to the EGFP fusion construct used for SIM, paGFP was fused to *E. coli* $F_oF_1$-ATP synthase subunit *a*, but encoded by plasmid pBH107 in *E. coli* strain RA1. In contrast, mEOS3.2 fused C-terminally to subunit *a* of $F_oF_1$-ATP synthase was integrated into the bacterial chromosome to ensure the appropriate, normal expression level of $F_oF_1$-ATP synthases present in each cell.

Figure 3 A and B show 3D-PALM images of paGFP-$F_oF_1$-ATP synthases in *E. coli* cells at 22°C. In Fig. 3 A, 204 $F_oF_1$-ATP synthases were localized at mid height in this cell using the localization information of the z position. By limiting z localizations to a range of 100 nm around a selected z position, the localization plane for the membranes was about 500 nm above the cover glass. Therefore, membranes were oriented perpendicular to the cover glass. In Fig. 3 B, a second cell from the same measurement is shown with 206 $F_oF_1$-ATP synthases localized at the mid height of the cell. Localization accuracies were distributed about 20 nm, as stated by the Nikon analysis software. Most of the membrane parts appeared to contain $F_oF_1$-ATP synthases evenly distributed, but in some membrane areas no fluorescent molecules could be localized.

For comparison, we studied $F_oF_1$-ATP synthase distribution also in the *E. coli* membrane parts that were oriented parallel to the cover glass. **Fehler! Verweisquelle konnte nicht gefunden werden.** show *E.coli* cells with localizations measured at 22°C (Figs. 3 C, D) and at 37°C (Figs. 3 E, F) using 2D- and 3D-PALM of mEOS3.2-$F_oF_1$-ATP synthases. 5000 images were recorded in the singe enzyme tracking SPT mode (see below), *i.e.* the focal plane was placed on the bacterial membrane facing the cover glass and TIR illumination was used. In Fig. 3 C, 295 $F_oF_1$-ATP synthases were localized with an accuracy of ~16 nm. In Fig. 3 D, 665 ATP synthases were localized with an accuracy of ~17 nm. In Figs. 3 E and F measured at 37°C, 436 and 1004 ATP synthases were localized with accuracies of less than 20 nm, respectively. The cells in Figs. 3 C-E did not exhibit clear patches of localized enzymes, but in Fig. 3 F a patterned distribution of $F_oF_1$-ATP synthases seemed to be present in the cytoplasmic membrane measured at 37°C.

In each case, the cells were grown in minimal medium M9 with glucose at 37°C, whereas the temperature for the PALM measurements was adjusted to 22°C, 30°C or 37°C on the microscope. Therefore, to further evaluate an influence of the measurement temperature on the distribution of $F_oF_1$-ATP synthases in *E. coli* membranes, we increased laser intensities and analyzed longer image sequences. Resulting 3D-PALM images of mEOS3.2-$F_oF_1$-ATP synthase in cytoplasmic membranes of *E. coli* are shown in Figure 4, measured at 22°C (Fig. 4 A-C) and at 37°C (Fig. 4 D-E), respectively. For the three cells measured at 22°C (Fig. 4 A-C), enzymes were selected that were localized in the z range between -400 nm and +400 nm around the focal plane set to 500 nm above the cover glass. 1304 (A), 1157 (B) and 798 (C) $F_oF_1$-ATP synthases were localized using photon count thresholds between 500 to 2000 photons per spot, resulting in an average accuracy between 48 and 56 nm according to the Nikon software.

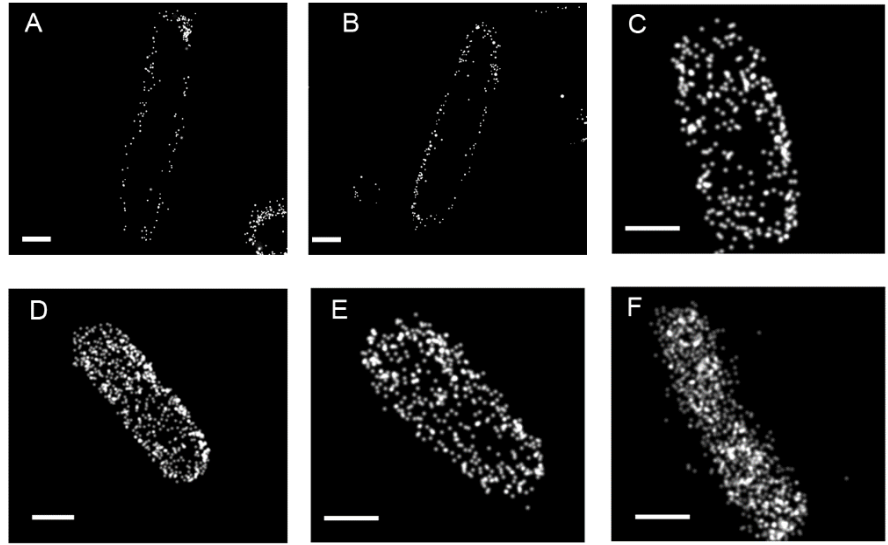

**Figure 3: A, B**, 3D-PALM localizations of $F_oF_1$-ATP synthases in living *E. coli* cells (RA1/pBH107) at 22°C using paGFP fused to subunit *a*. **C-F,** 2D-PALM localizations of mEOS3.2 fused to $F_oF_1$-ATP synthase (*E. coli* HW3) using TIR-illumination at 22°C (**C, D**) and at 37°C (**E, F**), respectively. Horizontal scale bars indicate 500 nm in length.

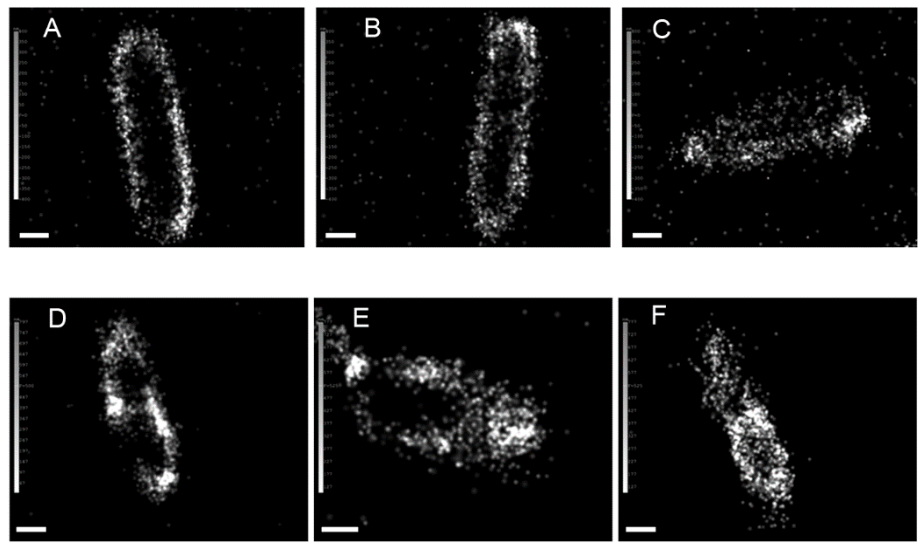

**Figure 4**: **A-C**, 2D-visualization of 3D-PALM localizations of mEOS3.2 fused to subunit *a* of $F_oF_1$-ATP synthase (*E. coli* HW3) at 22°C. **D-F**, 2D-visualization of 3D-PALM localizations of mEOS3.2 fused to subunit *a* of $F_oF_1$-ATP synthase (*E. coli* HW3) at 37°C. Horizontal scale bars indicate 500 nm length (x and y), vertical scale bars indicate gray-scale coding for z positions in each image.

The same type of 3D-PALM measurements were then carried out at 37°C. Fig. 4 D-F shows the localization images. Mean accuracies were determined to ~16 nm, and thresholds for localization were set to photon counts between 500 and up to 6000. A total number of 9.618 $F_oF_1$-ATP synthases were found in the region of interest. 927 to 1149 enzymes were localized in Fig. 4 D-F with maximal accuracies ranging from 41 to 47 nm. Although the number of localized $F_oF_1$-ATP synthases per cell and the localization accuracy were found similar for measurements at 22°C and at 37°C, the $F_oF_1$-ATP synthases seemed to be localized more unequally distributed over the cytoplasmic membrane at 37°C. This might be attributed to several $F_oF_1$-ATP synthases in close vicinity, eventually reinforced by a maximal photon count threshold that was set three times higher than for localizations at 22°C.

## 3.4 Diffusion analysis by single particle tracking

For SPT measurements 400 cells in total were analyzed, yielding more than 60,000 tracks of single diffusing $F_oF_1$-ATP synthases in *E. coli* membranes. We proceeded as follows: first, EMCCD images in the Nikon data format 'nd2 files' were converted to 'tif' stacks using the 'Bio-Formats plugin' for ImageJ to conserve the full bit depth. We then analyzed the 'tif' data in two different ways. For a more general, but preliminary analysis we used the 'trackmate plugin' in ImageJ to identify molecules, localized them in each image and built tracks for the same molecule. The 'trackmate plugin' settings used for particle detection were (*i*) 'DoG Detector with 0.3 µm blob diameter', (*ii*) '6.0 threshold', and (*iii*) 'sub-pixel localization enabled'. Tracks were built by a nearest neighbor search with a maximum distance of 0.2 µm (*i.e.* a maximum diffusion length between individual images or frames). Found tracks were exported as 'xml files' for further analysis in Matlab using the 'MSD Analyzer package'. An overview of found tracks in a single *E. coli* cell is shown in Fig. 5 A, and three examples revealing different diffusion behavior of $F_oF_1$-ATP synthases are shown in Fig. 5 B.

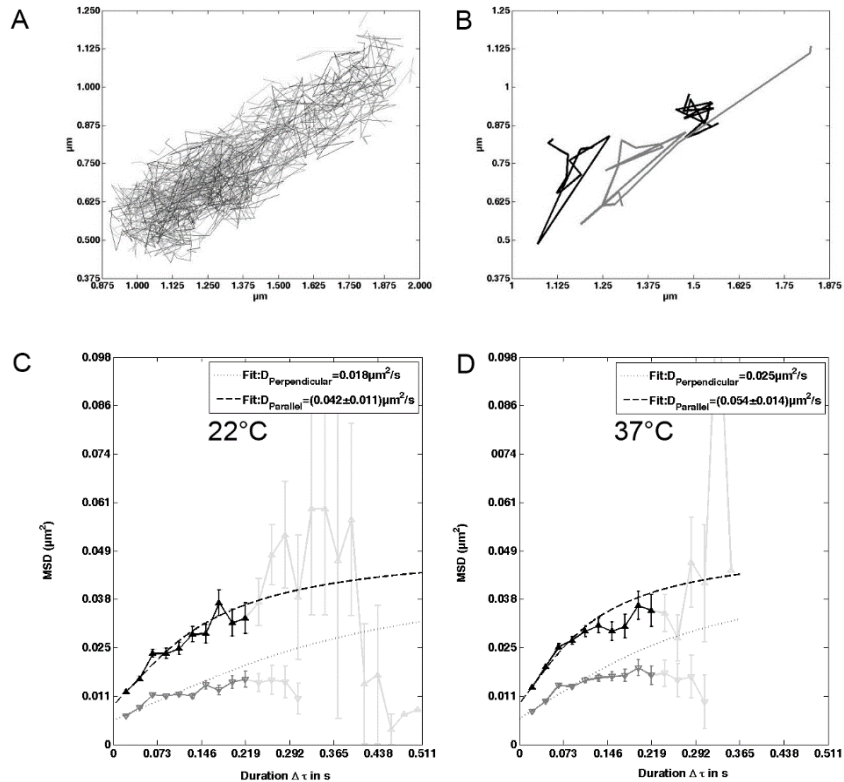

**Figure 5**: **A, B**, Visualization of individual tracks of mEOS3.2-$F_oF_1$-ATP synthase (*E. coli* HW3) found by the 'trackmate plugin' for ImageJ in a single living *E. coli* cell. (**A**) 280 tracks were found in this single cell; (**B**), three exemplary tracks with different diffusion properties. **C, D,** Diffusion analyses for all tracks at 22°C (**C**) and at 37°C (**D**) separated into diffusion components parallel to the long *E. coli* axis as well as to the perpendicular, short axis; see text. Error bars represented standard deviation of the mean.

'Trackmate' does not take into consideration the curvature of the *E. coli* membrane, *i.e.* the distortion of tracks caused by imaging a three-dimensional cylindrical surface projected to a two-dimensional imaging plane. However, 'trackmate' allowed to look at each track separately and to look at the distribution of MSDs and thus distinct diffusion speeds for individual $F_oF_1$-ATP synthases (data not shown). For a more in depth analysis we used our previously developed Matlab code as described (5). This analysis takes into account the shape of the *E. coli* membrane, therefore, correcting for perspective distortion. Furthermore, it separates diffusion into two components, perpendicular and parallel to the long cell axis. This is necessary because of the limited size of the *E. coli* cell and the corresponding statistical pitfalls for diffusion analysis.

In TIR illumination mode, the observable part of the cytoplasmic cell membrane is limited to areas near the cover glass/bacterium interface. Thereby background fluorescence is minimized and imaging single fluorescent proteins achieves high SNR contrast. Correcting for *E. coli* membrane geometry would still lead to a singularity at the parts of the cell membranes that are perpendicular to the image plane. However, these membrane parts are further away from the surface and get only weakly excited by TIR. Both effects limit the useful observable area to a size smaller than the diameter of the cell. Despite the fact that we did not account for the specific imaging problems of motion blur or EMCCD noise properties, average diffusion parameters of $F_oF_1$-ATP synthase in *E. coli* membranes could be fitted with the previous formula(5) and temperature influences could be compared. Fig. 5 C (measured at 22°C) and Fig. 5 D (measured at 37°C) show the mean-squared-displacement (MSD) fittings. Error bars represented the standard deviation of the mean. The 2D SPT tracks were split into diffusion components perpendicular to the cellular axis. Dashed or dotted lines were the fits to our diffusion equation to the MSD. These fittings represented the behavior for averaged 'free' diffusion, taking into account only the apparent confinement by the limited observation area in the *E. coli* membrane.

We found a mean longitudinal diffusion coefficient $D_{ll}$ = (0.042 ± 0.011) µm$^2$/s for mEOS3.2-$F_oF_1$-ATP synthase at 22°C, $D_{ll}$ = (0.084 ± 0.036) µm$^2$/s at 30°C (data no shown), and $D_{ll}$ = (0.054 ± 0.014) µm$^2$/s at 37°C.

## 4. DISCUSSION

$F_oF_1$-ATP synthases were labeled for live cell imaging in the cytoplasmic membranes of *E. coli* bacteria. Genetic fusions of three different fluorescent proteins EGFP, paGFP or mEOS3.2 to the C terminus of subunit *a* in the membrane-embedded $F_o$ part were achieved and compared. Enzymes were expressed in *E. coli* strain RA1 encoded on a plasmid (pSD166 with EGFP, pBH107 with paGFP), or chromosomally encoded in the new *E. coli* strain HW3 (with mEOS3.2). Labeled $F_oF_1$-ATP synthase expression levels were compared and analyzed with respect to 'wild type' enzymes without a protein fusion to subunit *a*. For all three mutants, the expression of the expression of the $F_oF_1$ complexes was lowered to about 50% according to immunoblotting with antibodies raised against subunits of $F_oF_1$ and Coomassie Blue staining of membrane proteins in SDS-PAGE. However, ATP hydrolysis activities of mEOS3.2-$F_oF_1$-ATP synthase containing membranes were found similar to membranes of the same *E. coli* strain with the 'wild type' enzymes. We conclude that the bulky fusion proteins (additional 29 kDa) did not disturb the function of the enzymes, but have only an impact on the expression due to the insertion of the fusion gene into the *atp* operon. Activity results have been reported previously for the EGFP fusion to $F_oF_1$-ATP synthase encoded by plasmid pSD166 or pDC61(23, 26, 34-37) as well as for the same mutant but chromosomally encoded in *E. coli* strain BW25113(9, 11).

Confocal imaging(11) and localization microscopy(9) of chromosomally encoded EGFP-$F_oF_1$-ATP synthases in *E. coli* strain BW25113 provided evidence that these membrane enzymes are confined to patches with sizes between 70 nm and 250 nm depending on the optical resolution limit of the microscopy approach. Therefore, we re-investigated the distribution of plasmid encoded enzymes in *E. coli* strain RA1. Structured illumination microscopy with a Nikon N-SIM setup revealed that the enzymes are clearly embedded in the cytoplasmic membrane, and no freely diffusing EGFP was detected within the cytosol. The distribution of EGFP-$F_oF_1$-ATP synthases appeared to be mostly homogeneous in the membrane when measured at 22°C. However, some brighter membrane spots were also found in a few cells. Spatial and temporal distribution of these brighter spots or possibly microcompartments were not quantified, because SIM only provides a resolution improvement down to about 100 nm, which is comparable to ten times the diameter of a single $F_1$ part of the enzyme.

In our previous *E. coli* imaging approach(5), we photobleached the EGFP-$F_oF_1$-ATP synthases until we reached the level of only few remaining fluorescent enzymes in the membrane. As photobleaching with high laser power might also damage other enzymes, change the diffusion of proteins in the membrane and, furthermore, and could influence the metabolism and the life cycle of the *E. coli* bacteria in general, we use photoactivation of non-fluorescent paGFP-$F_oF_1$-ATP synthases or photoswitching of non-detectable mEOS3.2-$F_oF_1$-ATP synthases for PALM imaging in this study. The Nikon N-STORM setup allows us to use different laser wavelengths simultaneously as well as sequentially, so that specific activation of the fusion proteins can be discriminated from non-specific fluorescence of single molecules. The obtained 2D and 3D localization images of $F_oF_1$-ATP synthases indicated that these mutant enzymes are embedded in the cytoplasmic membranes as well. The distributions of localized enzymes in the membranes, *i.e.* in the membrane parts perpendicular to the cover glass or in the horizontal parts excited by TIR, did not provide clear evidence for enzyme patches when measured at 22°C. However, apparent patches or non-homogeneous distributions were found in many *E. coli* cells measured at 37°C.

It is still not clear to us what could have caused these distributions at higher temperatures. We noted that up to 20 min passed between attaching the *E. coli* cells in our biochemistry lab to the poly-lysine coated cover glass, sealing the sample chambers with nail polish and transporting the samples at 37°C to the microscope. This might affect not only the growth of *E. coli* in the sample chamber resulting in fast consumption of $O_2$, but could also initiate the catabolic removal of unnecessary OXPHOS complexes like the $F_oF_1$-ATP synthases in the absence of $O_2$ as terminal electron acceptor for the respiratory chain. Furthermore, the proton motive force might be rapidly dissipated under these conditions and it has been shown recently that at least the (trans)membrane potential modulates the localization of integral membrane proteins in the cytoplasmic membrane(38). On the other hand, single-molecule *in vivo* imaging revealed that the bacterial respiratory complexes have a delocalized oxidative phosphorylation system, however, it includes independent patches of each of the different functional OXPHOS complexes involved, and rapid, long-range diffusion of ubiquinone is used to shuttle the electrons between the complexes involved(9). Nevertheless, for future experiments, temperature control for imaging on the microscope has to be accompanied by microfluidic control and exchange of the media around the cells. With a life cycle of about 20 minutes, *E. coli* cells have to be imaged on longer time scales to identify the viable and active cells and to discard aging or stalling cells.

The photoswitchable fluorescent protein mEOS3.2 fused to $F_oF_1$-ATP synthases could also be used for diffusion analysis of the membrane enzymes. Thereby active photobleaching like in the case of EGFP could be avoided. The resulting one-dimensional diffusion coefficients indicated a slightly faster diffusion in the membranes at 37°C than at 22°C, and were comparable to the one-dimensional diffusion coefficient of EGFP-fused $F_oF_1$-ATP synthase D = (0.072±0.015) µm$^2$/s at 25°C that we have measured previously with a different microscope(5), but using the same diffusion model for confined diffusion due to the limited elongation of *E. coli* along the short axis. However, other diffusion models for bacterial membrane proteins exist which take the specific measurement conditions with EMCCD cameras into account, *i.e.* localization precision and motion blur, and allow to extract not only a mean D value, but also contributions of several differently diffusing molecules(9, 39, 40). In future work, we will compare these available diffusion models with our approach.

Evaluating EGFP and the photoactivatable proteins as labels for single $F_oF_1$-ATP synthases is directly related to our current research. In part we focus on biophysical measurements of conformational changes of the *E. coli* enzyme by single-molecule FRET(41-58) (<u>F</u>örster <u>r</u>esonance <u>e</u>nergy <u>t</u>ransfer*) in vitro* as well as other ATP-driven membrane transporters like KdpFABC(59) and Pgp(60, 61). The next research steps will include similar measurements in living *E. coli* cells, and, therefore, new labeling strategies based on specifically attached fluorescent protein have to be employed. Due to the small size of *E. coli* as seen in SIM and PALM images above, and an estimated number of hundreds to thousands of $F_oF_1$-ATP synthases in the cytoplasmic membrane, a simple photobleaching approach of EGFP as a FRET donor label will not work. Instead, controlled photoactivation of the most photostable fluorescent protein as a fusion could become a reliable alternative.

The other current research focus is the spatio-temporal characterization of the assembly process of the membrane protein $F_oF_1$-ATP synthase with its 22 subunits(12, 22, 27, 28, 62-74). Here, fusion of photoactivatable fluorescent protein to different subunits of the membrane-embedded $F_o$ part or the soluble $F_1$ part with different spectral properties might become a tool to study these processes in living *E. coli* cells in real time and at the level of single functional nanomachines.

**Acknowledgements**

We thank Dr. Pingyong Xu (Beijing, China) and Jakob Piehler (Osnabrück, Germany) for kindly providing plasmids pmEos3.2-N1 and pSems-paGFP, respectively, Brigitte Herkenhoff-Hesselmann (Osnabrück, Germany) and Sonja Rabe (Jena, Germany) for expert technical assistance, and Thorsten Rendler (Stuttgart, Germany) for providing the Matlab scripts to analyze diffusion. This work was supported in part by DFG grants BO1891/10-2 (to M.B.), DE482/1-2 (to G.D.H.) and by the Sonderforschungsbereich 944 ‚Physiologie und Dynamik zellulärer Mikrokompartimente' (to G.D.H.). The Nikon N-SIM / N-STORM microscope was funded by the state of Thuringa (grant FKZ 12026-515 to M.B.).